\documentclass[conference]{IEEEtran}
\IEEEoverridecommandlockouts
\usepackage{cite}
\usepackage{amsmath,amssymb,amsfonts}
\usepackage{algorithmic}
\usepackage{graphicx}
\usepackage{textcomp}
\usepackage{xcolor}
\def\BibTeX{{\rm B\kern-.05em{\sc i\kern-.025em b}\kern-.08em
    T\kern-.1667em\lower.7ex\hbox{E}\kern-.125emX}}
\usepackage{amsmath, amsfonts, amssymb,graphicx}
\usepackage{booktabs} 
\usepackage{multirow} 
\usepackage{cite} 
\usepackage{tabularx}
\usepackage{stackengine}

\usepackage{array}
\usepackage{xcolor}
\usepackage{threeparttable}
\usepackage{pgffor}
\usepackage{tikz}
\usetikzlibrary{calc}
\usepackage{tikz,pgfplots}
\usepackage{tikz-3dplot}
\usepackage{hyperref}
\usepackage{cleveref}
\usepackage{caption}
\usepackage[compatibility=false]{caption}
\usepackage{subcaption}
\usepackage{url} 
 \usetikzlibrary{shapes.geometric, arrows}
\usetikzlibrary{patterns}
\usetikzlibrary{intersections}
\usepackage{CJKutf8}
\usepackage{pgfplots}
\usetikzlibrary{shadings,intersections,quotes,angles}
\usepackage{verbatim}
\usetikzlibrary{calc,fadings,decorations.pathreplacing,intersections}
\pgfplotsset{compat=1.18}
\usepackage{lettrine}
\usepackage{stackengine}

\graphicspath{{figures/}}

\usepackage{cleveref}
\tdplotsetmaincoords{70}{120}

\definecolor{myred}{RGB}{0,0,0}
\begin{document}

\title{Data-independent Beamforming  for \\ End-to-end Multichannel Multi-speaker ASR
}

\author{\IEEEauthorblockN{
Can Cui\thanks{This work was mostly conducted while Can Cui was pursuing her PhD with Inria Centre at Université de Lorraine (Nancy, France).}}
\IEEEauthorblockA{\textit{Research and Development Group} \\
\textit{iFLYTEK Co., Ltd.}\\
Shanghai, China\\
cancui11@iflytek.com}
\and
\IEEEauthorblockN{
Paul Magron,
Mostafa Sadeghi,
Emmanuel Vincent }
\IEEEauthorblockA{\textit{Multispeech} \\
\textit{Université de Lorraine, CNRS, Inria, LORIA, F-54000} \\
Nancy, France \\
\{paul.magron, mostafa.sadeghi, emmanuel.vincent\}@inria.fr
}}

\maketitle

\begin{abstract}
Automatic speech recognition (ASR) in multichannel, multi-speaker scenarios remains challenging due to ambient noise, reverberation and overlapping speakers. In this paper, we propose a beamforming approach that processes specific angular sectors based on their spherical polar coordinates before applying an end-to-end multichannel, multi-speaker ASR system. This method is data-independent and training-free. We demonstrate that using a group of beamformed signals improves ASR performance compared to using the same number of raw microphone signals. Moreover, increasing the number of signals used for beamforming further enhances recognition accuracy, leading to a more efficient use of multichannel signals while reducing the overall input load for the ASR system. We conduct experiments on the AMI meeting corpus, where the proposed method reduces word error rate by up to 11\% and improves speaker counting accuracy by up to 27\% relative compared to a multichannel ASR baseline system that does not exploit beamforming.
\end{abstract}

\begin{IEEEkeywords}
multichannel multi-speaker speech recognition, data-independent beamforming, spatial information
\end{IEEEkeywords}

\section{Introduction}

Transcribing real-world distant-microphone multi-speaker meetings recordings is an active research area~\cite{yu2022m2met,cornell2023chime,cornell2024chime}. However, it remains a challenging task due to noise, reverberation, and overlapping speech. To address these challenges, numerous studies~\cite{chang21_interspeech,li2022pcg,yu2023mfcca,shi2023comparative,cui2023} have explored end-to-end multichannel architectures. Multichannel audio processing introduces a wealth of spatial information, making it a valuable approach for speech separation and enhancement. By utilizing multiple microphones, spatial cues from various directions can be harnessed to improve the overall robustness and performance of these systems.
Inspired by single-channel attention~\cite{vaswani2017attention}, the multi-frame cross-channel attention (MFCCA) mechanism~\cite{yu2023mfcca} either concentrates on understanding global correlations between sequences across different channels or efficiently utilizes detailed channel-wise information at each time step.
This mechanism can then be integrated within an ASR encoder to effectively capture spatial information from multiple microphones,
where it has been shown to outperform single-channel ASR systems~\cite{yu2023mfcca,cui2023}. However, it directly processes raw far-field microphone signals, making it susceptible to ambient noise and reverberation, which can degrade recognition performance. 


Several studies have been conducted to enhance multichannel input quality, particularly through beamforming \cite{zhang2022all,masuyama2023end,cui2024joint} and/or speech separation \cite{raj2021integration,shi2022train,cornell2023chime}. While these methods improve speech quality, they typically rely on subsequent single-channel ASR architectures, leading to the loss of spatial information during recognition.
To mitigate this, \cite{kanda2023vararray} proposes feeding a multichannel ASR system with beamformed outputs as separate input channels, while \cite{mu2024automatic} explores selecting a subset of audio channels that are optimal for the task at hand. However, these approaches require data-dependent beamforming and additional training, making the process more complex.

In this paper, we alleviate the afore-mentioned issues by leveraging a beamforming approach that extracts signals from specific angular sectors. This allows to enhance speech quality by capturing speech information from various participant positions. This multichannel input is then fed into an end-to-end ASR system based on serialized output training (SOT)~\cite{kanda2020serialized} within a multichannel architecture using the MFCCA mechanism to further exploit spatial information. Our proposed method is data-independent and does not require training for beamforming, which makes it adaptable to any multichannel ASR system.
We conduct experiments on the real-world AMI meeting dataset~\cite{carletta2005ami}.
Our results show that the proposed approach reduces word error rate (WER) by up to 11~\% relative and speaker counting accuracy by up to 27~\% relative, when compared to an ASR that uses raw multichannel signals as input. Our system also reduces WER by 10~\% relative to an MVDR-based single-channel ASR baseline.

The rest of this paper is organized as follows. Section~\ref{sec:methods} presents the proposed data-independent beamforming technique. We then describe our experimental setup and results in Section~\ref{sec:exp}. Finally, Section~\ref{sec:conclusion} concludes the paper.

\begin{figure*}[!t]
  \centering
  \includegraphics[width=0.68\linewidth]{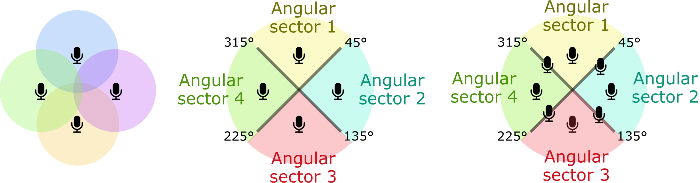}
  \caption{Bird's-eye view of the acoustic scene for original microphone channels (left), and beamformed channels corresponding to 4 angular sectors from either 4 microphones (center) or 8 microphones (right).}
  \label{fig:angles}
\end{figure*}
  

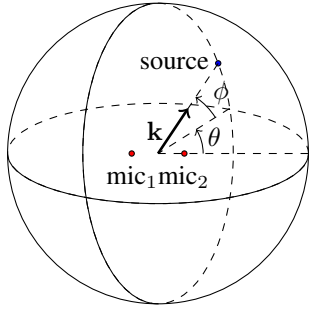
\begin{figure}[!tbp]
    \centering
\begin{tikzpicture}[scale=0.5]
  \def\R{4} 

  \coordinate (O) at (0,0);
  \coordinate (mic1) at (-0.7,0);
  \coordinate (mic2) at (0.7,0);
  \coordinate (source) at (\R/2.5, 0.6*\R, 0);
  \coordinate (a) at (\R/2.1, 0.3*\R, 0);
\coordinate (b) at (\R,0);
  \filldraw[fill=red] (mic1) circle (2pt) node[below, left=-3ex, yshift=-2ex] {mic$_1$};
  \filldraw[fill=red] (mic2) circle (2pt) node[below, right=-3ex, yshift=-2ex] {mic$_2$};
  \filldraw[fill=blue] (source) circle (2pt) node[left] {source};

\draw[dashed] (source) -- (O);
\draw[->,line width=1pt] (O) -- ($(source)!0.5!(O)$) node[midway, above, left] {$\mathbf{k}$};

\draw[dashed] (a) -- (O);
\draw[dashed] (b) -- (O);

  \draw (O) circle (\R);
  \draw[dashed] (O) ellipse (\R{} and \R/3);
  
    \draw[dashed] (O) ellipse (\R{}/2 and \R{});
    
    \begin{scope}
    \clip (O) -- (90:\R) arc (90:270:\R) -- cycle;
    \draw(O) ellipse (\R{}/2 and \R{});
\end{scope}

\begin{scope}
    \clip (O) -- ++(180:\R) arc (-180:0:\R) -- cycle;
    \draw(O) ellipse (\R{} and \R/3);
\end{scope}
    
\pic[draw,->,angle radius=0.6cm,angle eccentricity=1.3,"$\theta$"] {angle=b--O--a};

\pic[draw,->,angle radius=0.9cm,angle eccentricity=1.3,"$\phi$"] {angle=a--O--source};

\end{tikzpicture}
\caption{Geometrical illustration of the azimuth~$\theta$, the elevation~$\phi$ and the unit vector $\textbf{k}$ between the source and the microphone array center.}
\label{fig:sphere}
\end{figure}

\section{Data-independent beamforming based on target angular sectors}
\label{sec:methods}

We propose to preprocess the multichannel signals with data-independent beamformers, so that each resulting beamformed signal focuses on sound incoming from a specific angular sector. This preprocessing makes it possible to exploit the phase information corresponding to
distinct speaker positions. The beamformers for each angular sector are calculated from all microphones, so the number of microphones and the number of angular sectors are two independent concepts, as illustrated in \Cref{fig:angles}. Let $I$ and $S$ denote the number of microphones and angular sectors, respectively. Multichannel signals in the short-time Fourier transform (STFT) domain are denoted $\textbf{x}(f) \in \mathbb{C}^{I}$ at frequency $f$, as we discard the time frame index for clarity.

\subsection{Data-independent beamforming}

We consider the data-independent beamformer presented in \cite[Chap.\-10.3]{vincent2018audio}, which is designed to extract signals from specific angles by leveraging prior knowledge about the speakers' positions. Assuming that the sources are in the far field of the array \cite{don1993array}, the problem can be defined as finding the beamformer $\textbf{w}(f) \in \mathbb{C}^{I}$, whose spatial response
is closest to a predefined target $b^\text{tgt}(\theta, \phi, f)$:
\begin{multline}
\operatorname*{argmin}_\textbf{w} 
\frac{1}{4\pi} \int_{\Omega} |\textbf{w}^H\textbf{d}(\theta, \phi, f) - b^\text{tgt}(\theta, \phi, f)|^2 
\cos\theta d\theta d\phi,
\label{eq:beamformer}
\end{multline}
\noindent where $\Omega$ denotes the set of all azimuth $\theta$ and elevation $\phi$ angles (these are illustrated in \Cref{fig:sphere}), $.^H$ denotes the Hermitian transpose,
and:
\begin{align}\label{eq:eq4}
\textbf{d}(\theta, \phi, f) &= \begin{bmatrix}  e^{-2\jmath \pi \textbf{k}^T(\theta, \phi, f)\textbf{m}_1/\lambda} \\
\vdots \\
e^{-2\jmath \pi \textbf{k}^T(\theta, \phi, f)\textbf{m}_I/\lambda} 
\end{bmatrix}
\end{align}
is the steering vector with $\textbf{m}_i$ the Cartesian coordinates of microphone $i$, $\lambda$ is the center wavelength of the narrowband signal, and:
\begin{align}\label{eq:eq5}
\textbf{k} &= \begin{bmatrix}  \cos\theta \cos\phi \\
\sin\theta \cos\phi \\
\sin\phi
\end{bmatrix}
\end{align}
is the unit vector from the center of the microphone array to the target source.

Since~\Cref{eq:beamformer} has no general closed-form solution, several solutions have been proposed for specific array geometries, predefined target models $b^\text{tgt}$, and/or exploiting numerical approximation schemes (see~\cite[Chap.\-10.3]{vincent2018audio} for an overview). In what follows we propose a simple predefined target model for $b^\text{tgt}$ which allows to derive a closed-form beamformer for each angular sector.

\begin{figure}[!tbp]
  \centering
  \includegraphics[width=0.58\linewidth]{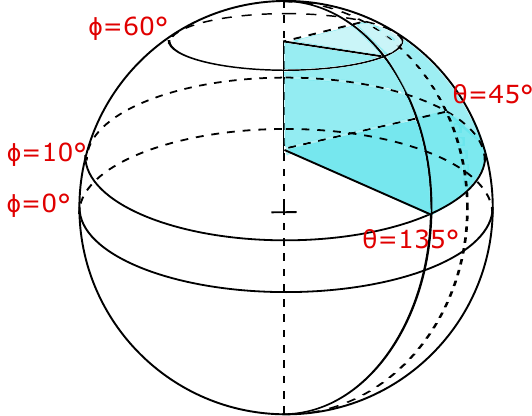}
  \caption{Angular sector corresponding to $\theta \in [45^\circ, 135^\circ]$ and $\phi \in [10^\circ, 60^\circ]$.}

  \label{fig:sphere_angu}
\end{figure}

\begin{figure*}[!t]
  \centering
  \includegraphics[width=0.99\linewidth]{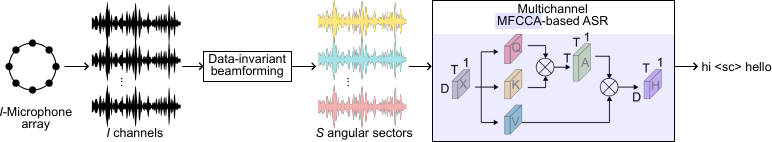}
  \caption{Overview of the proposed system combining data-invariant beamforming and multichannel MFCCA-based ASR. The MFCCA module extracts and integrates spatial-temporal features across the $S$ sectors, but can also directly process the $I$ raw multichannel signals.}
  \label{fig:syst}
  \vspace{-10pt}
\end{figure*}

\subsection{Proposed solution}
\label{subsec:solution}

Since the speech source is omnidirectional, we aim to extract a response from every spatial direction. Therefore, based on \Cref{eq:beamformer}, the beamformer  is obtained by integrating the signal over the responses from all azimuth and elevation angles. However, in this work we rather perform integration over \emph{angular sectors}, which represent groups of target position points in a given portion of space. Each angular sector $\Psi_s$, $s \in [1, S]$ is defined as a set of azimuth and elevation ranges, as illustrated in \Cref{fig:sphere_angu}.
We consider the following predefined target for each angular sector:
\begin{equation}
b_s^\text{tgt}(\theta, \phi, f) =
  \left\{
    \begin{aligned}
    & 1  \quad \text{if} \quad (\theta, \phi) \in \Psi_s, \\ 
    & 0  \quad \text{otherwise},
    \end{aligned}
    \right.
\end{equation}
which allows to derive the following closed-form solution for \Cref{eq:beamformer}:
\begin{multline}
\label{eq:solution1}
\hat{\textbf{w}}_s(f) =\left(\int_{\Omega} \cos\theta\times\textbf{d}(\theta, \phi, f)   \textbf{d}(\theta, \phi, f)^H d\theta  d\phi \right)^{-1} \\ \times \int_{\Psi_s}\cos\theta \times\textbf{d}(\theta, \phi, f)
 d\theta  d\phi.
\end{multline}

\noindent Once estimated, we apply the set of beamformers ${\hat{\textbf{W}}(f) \in \mathbb{C}^{I \times S}}$ to the input multichannel signal via $\hat{\textbf{W}_s}^H(f) \textbf{x}(f)$, and these are then reverted back to the time domain via inverse STFT.

Finally, the resulting beamformed signals are fed to a multichannel ASR system, as illustrated in \Cref{fig:syst}. We consider an MFCC-based ASR model, which captures angle-specific information and encodes dependencies across different spatial directions and time steps (more details in Section~\ref{subsec:systems}).


\subsection{Proposed angular sectors}
\label{subsec:beamform}

\begin{figure}[!t]
  \centering
  \begin{subfigure}[b]{0.95\linewidth}
    \centering
    \includegraphics[width=\linewidth]{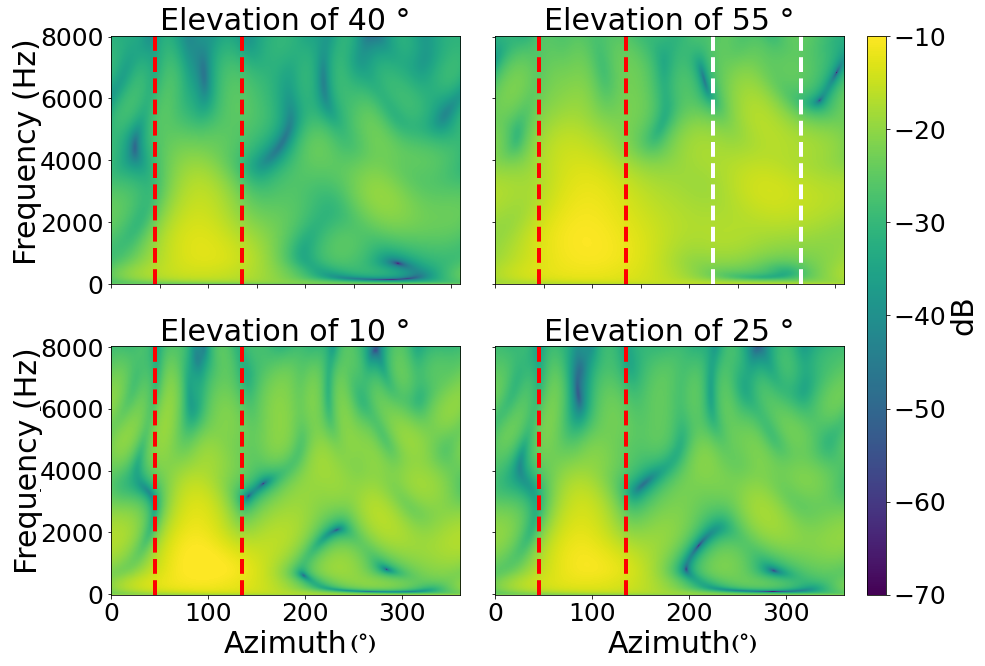}
    \caption{4 microphones}
    \label{fig:feature1}
  \end{subfigure}
  \hfill
  \begin{subfigure}[b]{0.95\linewidth}
    \centering
    \includegraphics[width=\linewidth]{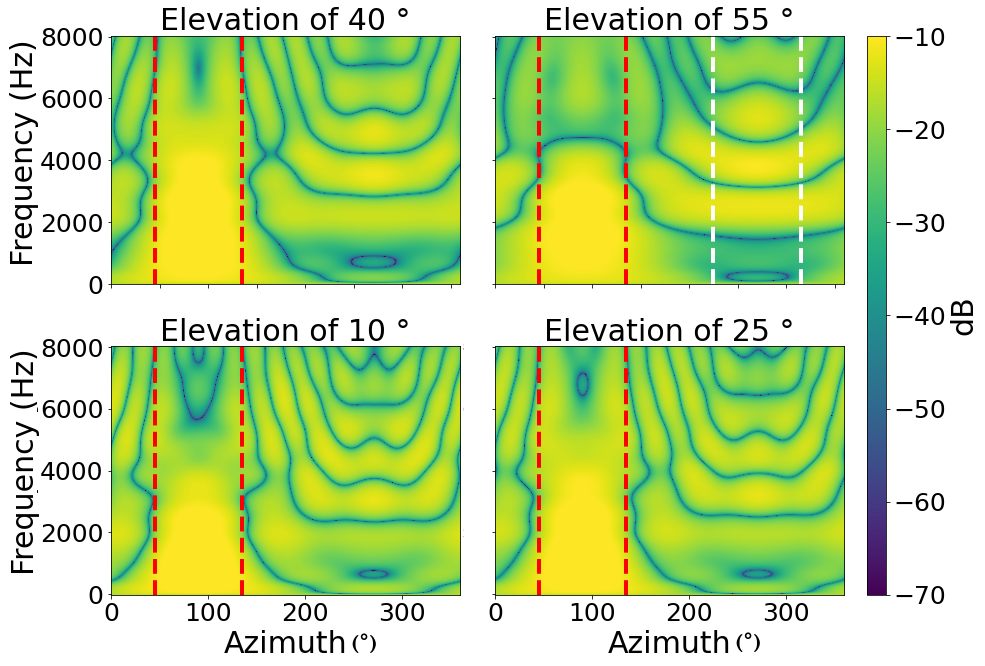}
    \caption{8 microphones}
    \label{fig:feature2}
  \end{subfigure}
  \caption [Response of the fixed beamformer associated with the second angular sector.]{Response of the fixed beamformer associated with the second angular sector. The dashed red and white lines define regions with azimuth values from 45 to 135°, and from 225 to 315°, respectively.}
  \vspace{-10pt}
  \label{fig:combined_features}

\end{figure}

In this paper, we consider four equally-spaced angular sectors ($S=4$), which corresponds to a four-speaker scenario as encountered in our dataset (see Section~\ref{subsec:data}). Note that this number can be adjusted to accommodate different meeting configurations. For realistic consideration of speakers' possible positions with respect to the microphone array, we choose an elevation range from 10 to 60 degrees for all sectors. The angular sectors cover a width of 90° azimuth each, ranging from 315 to 45°, from 45 to 135°, from 135 to 225°, and from 225 to 315°, respectively. While prior knowledge of the speaker's position is not essential for applying our data-independent beamformer, the beamformed signals are expected to isolate the target speakers given such a spatial setup. Besides, even though each beamformed signal does not perfectly extract each speaker, this preprocessing is expected to improve MFCCA's ability to model cross-channel dependencies.

We consider either 4 or 8 microphones arranged in a circular array, thus yielding three possible inputs to the multichannel ASR system, as illustrated in \Cref{fig:angles}: the original 4-channel signals from 4 microphones, the 4-channel beamformed signals extracted from 4 microphones, and the 4-channel beamformed signals extracted from 8 microphones.
\Cref{fig:combined_features} illustrates the response of the beamformer in the second angular sector, i.e., $\hat{\textbf{w}}_2^H(f) \textbf{d}(\theta, \phi, f)$, over frequencies and azimuth angles, for various elevation angles (10°, 25°, 40°, and 55°), and computed from either 4 or 8 microphones.
The 8-microphone setup maintains accurate performance (i.e., a high spatial response) up to 4,000 Hz, while the 4-microphone setup starts to degrade above 2,000 Hz.
The values within the target angular sector, an azimuth range from 45 to 135°, are higher than those in other sectors. For the 8-microphone setting, the quality of the response degrades above 4,000 Hz, especially at elevations of 10° and 55°, because they are closer to the border of the target elevation range, which is from 10° to 60°. At an elevation of 55° for both the 4- and 8-microphone settings, higher values can be found in other sectors, such as the fourth sector with an azimuth range from 225° to 315°. Overall, the 8-microphone responses are sharper and more accurate than those of the 4-microphone setting.


\section{Experiments}
\label{sec:exp}

This section details our experimental setup and results. For reproducibility, our code is available online.\footnote{\url{https://github.com/can-cui/data-invariant-beamforming-mc-asr}}

\subsection{Datasets and metrics}
\label{subsec:data}

\subsubsection{Datasets}
\label{subsubsec:data}
\textbf{Multi-speaker LibriSpeech} --- To achieve good performance on real-life meeting corpus, the ASR model must be pretrained on a larger simulated distant-microphone multi-speaker dataset \cite{yang2023simulating}. We created a 960~h training set and a 20~h development set from the LibriSpeech train-960 and dev-clean sets \cite{panayotov2015LibriSpeech}.
We adopted the room simulation settings as described in \cite{cui2023}.
Each mixture contains 1 to 3 speakers. 
Each speaker's utterance begins after a random delay of more than $0.5$~s from the start of the previous speaker's utterance, to better reflect realistic speaker turn-taking.
This also guarantees the first-in, first-out principle behind SOT. The sentences of each speaker are concatenated together and separated from the next speaker using a special \textless sc\textgreater\ token.\\[.75em]
\textbf{Real AMI} --- After it has been pretrained on Multi-speaker LibriSpeech, the ASR model is fine-tuned and evaluated on real AMI multiple distant microphone (MDM) data. We utilize the segmentation method in \cite{cui2023} to partition the MDM data into 5~s chunks and adjust the chunk start/end times to non-overlapped word boundaries. The resulting AMI dataset contains respectively 165~h, 19~h, and 19~h for training, development, and test. We consider 4- and 8-channel taken from Array1,\footnote{A few meetings have two arrays, each consisting of 8 microphones.} namely channels 1, 3, 5, and 7 in the 4-channel setting, and all 8 channels in the 8-channel setting.  We use the Full-corpus-ASR partition of the dataset.\footnote{See partition detail at \url{https://groups.inf.ed.ac.uk/ami/corpus/datasets.shtml}.}

For both datasets, audio signals are sampled at 16~kHz. Log-Mel spectrograms are extracted using a 25 ms window, a hop size of 10 ms, and 80 Mel bands. Note that even though in this study we consider static speakers and circular microphone arrays for simplicity, our proposed method can accommodate moving sources and any array geometry.

\subsubsection{Metrics}
\label{subsubsec:metrics}
We use the WER to evaluate the ASR task. The ASR model we employed transcribes speech from multiple speakers by inserting a speaker change token \textless sc\textgreater\ in a first-in, first-out order. This allows for calculating the number of speakers in both the ground truth and estimated transcriptions,
determined by counting the occurrences of \textless sc\textgreater\ tokens in the ASR output. As a result, we can also evaluate the speaker counting accuracy \cite{kanda2020joint} via the confusion scores:
\begin{align}\label{eq:confusion}
\text{Confusion\_score} (i, k) = \frac{N^i_{k}}{N_k},
\end{align}
where $N^i_{k}$ is the number of test signals where the estimate indicates $i$ speakers when the ground truth has $k$ speakers, and ${N_k}$ represents the number of test signals where the ground truth has $k$ speakers.

We employed the SCTK toolkit\footnote{\url{https://github.com/usnistgov/SCTK.git}} to conduct a Matched Pair Sentence Segment significance test. We indicate with bold fonts the best numerical results at a 0.05 significance level in the corresponding tables.

\subsection{Models and training setup}
\label{subsec:systems}

We implement our proposed beamformer as detailed in Section~\ref{sec:methods}. To compute it in practice, the integrals in \Cref{eq:solution1} are replaced with sums using a step size of 1 degree over the considered range, which allows for a good trade-off between computational burden and precision.

We then consider an MFCCA-based multichannel ASR model. It features 12 layers in the encoder and 6 layers in the decoder. The number of filterbank output channels and the number of bins in the estimated masks are both set at 513. All multi-head attention mechanisms have 4 heads,  the model dimension \(D \) is set to 256, and the size of the feedforward layer is 2,048. Following \cite{yu2023mfcca}, the context frame length \(F\) of MFCCA is set to 2. Our text tokenizer is a SentencePiece model \cite{kudo2018sentencepiece} with a vocabulary of 5,000 tokens.

To evaluate the effectiveness of the proposed beamforming method, as a comparison baseline we consider a neural MVDR beamformer \cite{lu2022towards,kim23b_interspeech}, as implemented in the torchaudio library~\cite{yang2021torchaudio}. This yields a single-channel signal which is then fed to a single-channel SOT-based ASR (see details above).

Note that even though more recent methods such as~\cite{kanda2020joint,kanda2023vararray} would be interesting comparison baselines, we could unfortunately not reproduce their results due to a lack of available code, data, and limited computational resources.

Our experiments were implemented using the SpeechBrain toolkit \cite{speechbrain}. Pretraining on LibriSpeech and fine-tuning on Real AMI are conducted using the Adam optimizer with a learning rate of $5\times 10^{-4}$ and $1\times 10^{-4}$, and a global batch size (batch size \(\times\) number of GPUs \(\times\) gradient accumulation factor) of 160 and 80, respectively.


\subsection{Evaluation results}

\begin{table*}[t]
\caption{ASR performance (WER in \%) on the LibriSpeech and Real AMI test sets, for several methods, according to the ASR input type (unprocessed, using the proposed beamforming, or an MVDR beamformer).}
\centering
\label{table:beam}
\scalebox{1}{
\addtolength{\tabcolsep}{-0.41em}
\begin{tabular}{c|c|c|c|c|c|c|c|c|c|cccccc}
    \hline
    \multirow{2}{*}{\bfseries Input type} &
    \multirow{2}{*}{\bfseries \# Microphones} & 
    \multirow{2}{*}{\bfseries \# Sectors} &
    \multicolumn{4}{c|}{\bfseries LibriSpeeh}&
    \multicolumn{4}{c}{\bfseries Real AMI}&
    \\  \cline{4-8} \cline{9-12} 
    &&&\textbf{1-speaker} & \textbf{2-speaker}& \textbf{3-speaker}& \textbf{1-3-speaker}&\textbf{1-speaker} & \textbf{2-speaker}& \textbf{3-speaker}& \textbf{1-4-speaker} \\
    \hline
    \multirow{1}{*} {Unprocessed}&\multirow{1}{*}{4 }& \multirow{1}{*}- &7.21&14.82&19.76&16.30&25.89&41.70&54.68&45.25 \\
   
    \hline
 \multirow{2}{*} {Proposed beamforming} &4 &\multirow{1}{*}4&7.46&12.27&22.99&17.64&24.52&\textbf{39.61}&52.19&43.14  \\
   &\multirow{1}{*}8  &\multirow{1}{*}4&\textbf{6.73}&\textbf{10.53}&\textbf{19.06}&\textbf{14.26}&\textbf{22.96}&40.01&\textbf{49.59}&\textbf{41.64} \\
  \hline
  
\multirow{2}{*} {MVDR} &\multirow{1}{*}4  &\multirow{1}{*}-&-&-&-& -&25.11&40.07&54.17&44.43 \\
 &\multirow{1}{*}8  &\multirow{1}{*}-&-& -&- & -&26.35&42.13&55.89&46.08 \\
    \hline
\end{tabular}
}
\vspace{-10pt}
\end{table*}

\subsubsection{Improvements on speech quality}

\begin{figure}[!tbp]
  \centering
  \includegraphics[width=.99\linewidth]{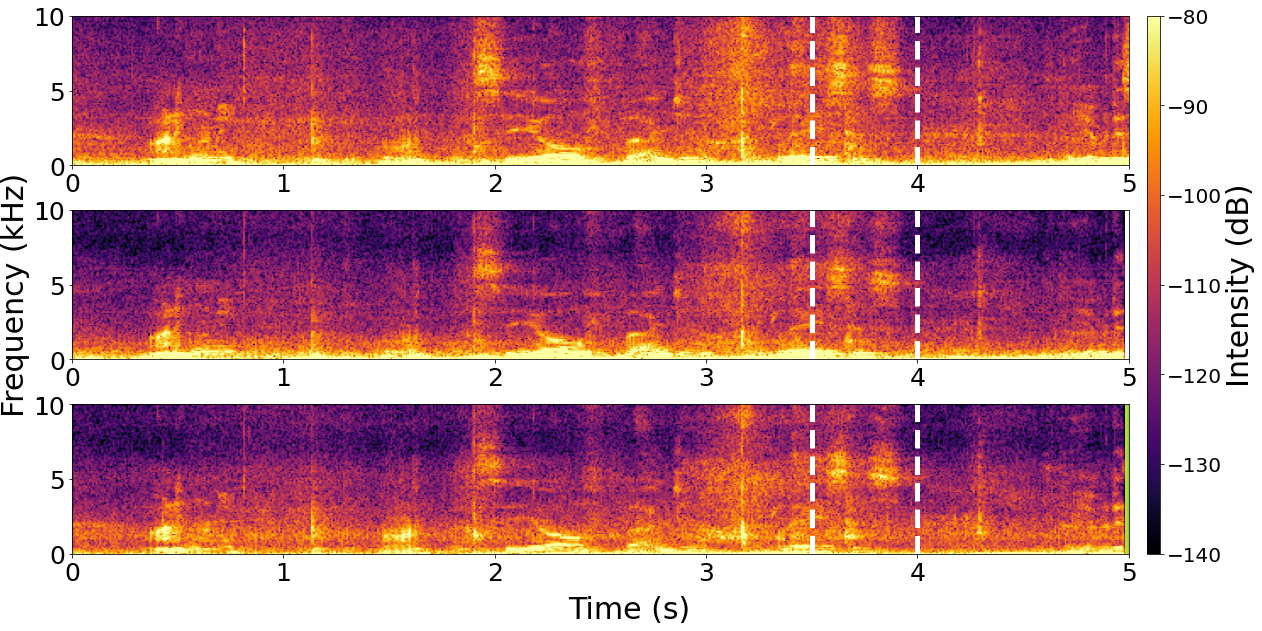}
  \caption [Spectrograms of one Real AMI test chunk.]{Spectrogram of one Real AMI test chunk: original 1$^{\text{st}}$ array channel (top), and beamformed signal in the 1$^{\text{st}}$ sector from 4 microphones (middle) or 8 microphones (bottom).
  }
  \label{fig:data-inv-sig}
\vspace{-10pt}
\end{figure}

First, we compare signals before and after beamforming by visualizing the spectrogram of a Real AMI meeting chunk in the first array / sector.
As shown in~\Cref{fig:data-inv-sig}, both beamformed signals (i.e., from either 4 or 8 microphones) significantly reduce ambient noise and reverberation compared to the original far-field channel. This illustrates the ability of the proposed beamformer to enhance the signal quality.
We also observe interfering speech reduction in the beamformed signal. Thus, even though the exact positions of the speakers are unknown, the proposed beamforming method is able to enhance a specific speaker (i.e., the one in the first angular sector) and reduce the other ones. Similar results are obtained from other arrays / sectors.

Furthermore, by comparing the signals extracted from 4 and 8 microphones, particularly in the zone between the white lines, we observe that the latter provides better dereverberation in the speech region by reducing interfering speech and enhancing the target speech. Similar phenomena can be observed in other speech segments, where the signal from 8 microphones shows clearer speech formants. This demonstrates that using more microphones
better exploits the available spatial information and thus yields cleaner speech signals.

\subsubsection{Speech recognition performance}

We then compare the results of ASR systems trained on original channels versus data-independent beamformed channels. \Cref{table:beam} displays the results on the LibriSpeech and Real AMI test sets using the pretrained and fine-tuned models, respectively. On the LibriSpeech test set, the model trained on 4-channel beamformed signals from 4 microphones (``beamformed-4") does not show improvement compared to the model trained on the unprocessed 4-channel signals, with WERs of 17.64\% vs. 16.30\% on the 1,2,3-speaker mixture test set, respectively. However, the model trained on 4-channel beamformed signals from 8 microphones (herein denoted ``beamformed-8") demonstrates a reduction of 13\% relative in WER compared to beamformed-4, reducing the WER from 16.30\% to 14.26\% on the 1,2,3-speaker mixture test set. For the 2-speaker mixture test set, both beamformed-4 and -8 show better performance compared to the unprocessed model, with a 14\% relative reduction in WER (from 14.82\% to 12.27\%) and a 29\% relative reduction in WER (from 14.82\% to 10.53\%), respectively.

On the Real AMI test set, both beamformed models yield a more substantial improvement compared to using the unprocessed signals. In particular, the beamformed-8 model achieves a reduction in  WER of 11\%, 4\%, 9\%, and 8\% relative compared to the unprocessed model for the 1, 2, 3, and {1-4}-speaker groups, respectively. Overall, these improvements highlight the efficiency of fine-tuning on beamformed signals to better exploit spatial information. Moreover, the beamformed-8 model consistently outperforms the beamformed-4 model on both the LibriSpeech and Real AMI test sets. This demonstrates that using beamformed signals from a larger number of microphones results in better performance for the MFCCA-vased multichannel ASR model. This conclusion is consistent with the fact that a larger number of microphones feeds more information to a given sector, thereby providing better enhancement of signals from that sector (see \Cref{fig:combined_features}).

\subsubsection{Speaker counting accuracy}

\begin{table}[!t]
\caption{Speaker counting accuracy (confusion score in \%) on the Real AMI test set for our proposed method.}
\label{table:spks-count-ami-chap3}
\centering
\scalebox{0.99}{
\addtolength{\tabcolsep}{-0.35em}
\begin{tabular}{c|c|c|c|c|c|cc}
\hline
    \multirow{2}{*}{\bfseries Input type} & 
    \multirow{2}{*}{\bfseries \# Speakers}& 
    \multicolumn{5}{c}{\bfseries Estimated \# speakers} & 
    \\ 
    \cline{3-7} 
    &&  1&2&3&4&\textgreater4
    \\ 
    \hline
     \multirow{4}{*} {Unprocessed} & 1 & 96.84 & 3.01& 0.14& 0.00 &0.00\\
     & 2  & 23.21 & \textbf{74.65}& 2.09& 0.03 &0.00 \\
     & 3  & 6.39 & 63.90& 29.08&0.61&0.00 \\
      & 4  & 2.85 &39.82& 52.60 & 4.71&0.00 \\
    \hline
    \multirow{4}{*} {Beamformed (8 mic.)} & 1 & \textbf{97.86} & 2.04& 0.09 & 0.00&0.00 \\
     & 2  & 25.89 & 72.05& 1.98 & 0.06&0.00\\
     & 3  & 6.29 & 56.71& \textbf{36.94}&0.04&0.00 \\
      & 4  & 1.86 & 33.87& 58.31& \textbf{5.70}&0.24 \\
\hline
\end{tabular}
}
\vspace{-10pt}
\end{table}

\Cref{table:spks-count-ami-chap3} presents the speaker counting accuracy on the Real AMI test set.\footnote{Note that these results are not directly comparable to similar studies utilizing speaker-attributed ASR \cite{cui2023,cui2024joint}, as these incorporate an additional speaker embedding model to enhance speaker prediction accuracy.} The results show that the model utilizing beamformed channels generally outperforms the one using original channels. Specifically, the proposed method improves speaker counting accuracy by 27\% relative (from 29.08\% to 39.94\%) in 3-speaker scenarios and by 21\% relative (from 4.71\% to 5.70\%) in 4-speaker scenarios. These findings highlight that beamforming improves the prediction of speaker change tokens by more effectively leveraging spatial information, particularly as the number of speakers increases.

\subsubsection{Comparison to MVDR-based beamforming}

We report the performance of the MVDR-based ASR method (see Section~\ref{subsec:systems}) in the last rows of~\Cref{table:beam}. We only consider test results on the AMI dataset, as it is more realistic. We observe that the model using raw multichannel input (i.e., no beamforming) does not yield a consistent improvement over the MVDR-ASR system. This suggests that the spatial information may not be fully exploited by the ASR model when it is not first processed by e.g., a beamformer. In contrast, when using our proposed data-invariant beamformed input, the performance improves significantly for both 4- and 8-channel configurations.
Notably, the model using 8 microphones achieves a 10\% relative reduction in WER, decreasing from 46.08\% in the MVDR-based system to 41.64\% on the mixed {1-4}-speaker scenarios. 
This result highlights the benefit of a robust, data-independent beamforming method that generalizes well across channel and speaker conditions, without relying on neural network-based learning.

\section{Conclusion}
\label{sec:conclusion}

In this paper, we have proposed a data-independent, training-free beamforming approach as input to an end-to-end multichannel multi-speaker ASR system. This beamformer captures speech from specific elevation and azimuth sectors, aligned with typical participant positions in meetings.
Our method achieves a reduced WER by up to 11~\% and improves speaker counting accuracy by up to 27~\% relative compared to using unprocessed multichannel signals, and also outperforms a competing method based on data-dependent MVDR beamforming. These results highlight the potential of data-invariant spatial filtering for efficient multi-speaker ASR. Future work will explore adaptive integration with downstream models, and deployment in more diverse acoustic conditions, including moving sources and ad-hoc microphone arrays.

\section{Acknowledgments}
\label{sec:acknowledgments}

Experiments presented in this paper were carried out using the \href{https://www.grid5000.fr}{Grid'5000}
testbed, supported by a scientific interest group hosted by Inria and including CNRS, RENATER and several Universities as well as other organizations.

\bibliographystyle{IEEEtran}
\bibliography{mybib}

\begin{thebibliography}{10}
\providecommand{\url}[1]{#1}
\csname url@samestyle\endcsname
\providecommand{\newblock}{\relax}
\providecommand{\bibinfo}[2]{#2}
\providecommand{\BIBentrySTDinterwordspacing}{\spaceskip=0pt\relax}
\providecommand{\BIBentryALTinterwordstretchfactor}{4}
\providecommand{\BIBentryALTinterwordspacing}{\spaceskip=\fontdimen2\font plus
\BIBentryALTinterwordstretchfactor\fontdimen3\font minus \fontdimen4\font\relax}
\providecommand{\BIBforeignlanguage}[2]{{%
\expandafter\ifx\csname l@#1\endcsname\relax
\typeout{** WARNING: IEEEtran.bst: No hyphenation pattern has been}%
\typeout{** loaded for the language `#1'. Using the pattern for}%
\typeout{** the default language instead.}%
\else
\language=\csname l@#1\endcsname
\fi
#2}}
\providecommand{\BIBdecl}{\relax}
\BIBdecl

\bibitem{yu2022m2met}
F.~Yu, S.~Zhang, Y.~Fu, L.~Xie, S.~Zheng, Z.~Du, W.~Huang, P.~Guo, Z.~Yan \emph{et~al.}, ``{M2Met: The ICASSP 2022 Multi-Channel Multi-Party Meeting Transcription Challenge},'' in \emph{Proc. of IEEE ICASSP}, 2022.

\bibitem{cornell2023chime}
S.~Cornell, M.~S. Wiesner, S.~Watanabe, D.~Raj, X.~Chang, P.~Garcia, Y.~Masuyam, Z.-Q. Wang, S.~Squartini, and S.~Khudanpur, ``The {CHiME-7 DASR Challenge}: Distant meeting transcription with multiple devices in diverse scenarios,'' in \emph{Proc. of CHiME}, 2023.

\bibitem{cornell2024chime}
S.~Cornell, T.~J. Park, H.~Huang, C.~Boeddeker, X.~Chang, M.~Maciejewski, M.~S. Wiesner, P.~Garcia, and S.~Watanabe, ``The {CHiME-8 DASR Challenge} for generalizable and array agnostic distant automatic speech recognition and diarization,'' in \emph{Proc. of CHiME}, 2024.

\bibitem{chang21_interspeech}
F.-J. Chang, M.~Radfar, A.~Mouchtaris, and M.~Omologo, ``Multi-channel transformer transducer for speech recognition,'' in \emph{Proc. of Interspeech}, 2021.

\bibitem{li2022pcg}
J.~Li, Y.~Zhu, D.~Luo, Y.~Liu, G.~Cui, and Z.~Li, ``The {PCG-AIID system for L3DAS22 challenge: MIMO and MISO} convolutional recurrent network for multi channel speech enhancement and speech recognition,'' in \emph{Proc. of IEEE ICASSP}, 2022.

\bibitem{yu2023mfcca}
F.~Yu, S.~Zhang, P.~Guo, Y.~Liang, Z.~Du, Y.~Lin, and L.~Xie, ``{MFCCA}: Multi-frame cross-channel attention for multi-speaker {ASR} in multi-party meeting scenario,'' in \emph{Proc. of IEEE SLT}, 2023.

\bibitem{shi2023comparative}
M.~Shi, J.~Zhang, Z.~Du, F.~Yu, Q.~Chen, S.~Zhang, and L.-R. Dai, ``A comparative study on multichannel speaker-attributed automatic speech recognition in multi-party meetings,'' in \emph{Proc. of APSIPA ASC}, 2023.

\bibitem{cui2023}
C.~Cui, I.~Sheikh, M.~Sadeghi, and E.~Vincent, ``End-to-end multichannel speaker-attributed {ASR}: Speaker guided decoder and input feature analysis,'' in \emph{Proc. of IEEE ASRU}, 2023.

\bibitem{vaswani2017attention}
A.~Vaswani, N.~Shazeer, N.~Parmar, J.~Uszkoreit, L.~Jones, A.~N. Gomez, {\L}.~Kaiser, and I.~Polosukhin, ``Attention is all you need,'' \emph{Advances in Neural Information Processing Systems}, vol.~30, 2017.

\bibitem{zhang2022all}
Z.~Zhang, T.~Yoshioka, N.~Kanda, Z.~Chen, X.~Wang, D.~Wang, and S.~E. Eskimez, ``All-neural beamformer for continuous speech separation,'' in \emph{Proc. of IEEE ICASSP}, 2022.

\bibitem{masuyama2023end}
Y.~Masuyama, X.~Chang, S.~Cornell, S.~Watanabe, and N.~Ono, ``End-to-end integration of speech recognition, dereverberation, beamforming, and self-supervised learning representation,'' in \emph{Proc. of IEEE SLT}, 2023.

\bibitem{cui2024joint}
C.~Cui, I.~A. Sheikh, M.~Sadeghi, and E.~Vincent, ``Joint beamforming and speaker-attributed {ASR} for real distant-microphone meeting transcription,'' in \emph{Proc. of EUSIPCO}, 2025.

\bibitem{raj2021integration}
D.~Raj, P.~Denisov, Z.~Chen, H.~Erdogan, Z.~Huang, M.~He, S.~Watanabe, J.~Du, T.~Yoshioka, Y.~Luo \emph{et~al.}, ``Integration of speech separation, diarization, and recognition for multi-speaker meetings: System description, comparison, and analysis,'' in \emph{Proc. of IEEE SLT}, 2021.

\bibitem{shi2022train}
J.~Shi, X.~Chang, S.~Watanabe, and B.~Xu, ``Train from scratch: Single-stage joint training of speech separation and recognition,'' \emph{Computer Speech \& Language}, vol.~76, p. 101387, 2022.

\bibitem{kanda2023vararray}
N.~Kanda, J.~Wu, X.~Wang, Z.~Chen, J.~Li, and T.~Yoshioka, ``Vararray meets {T-Sot}: Advancing the state of the art of streaming distant conversational speech recognition,'' in \emph{Proc. of IEEE ICASSP}, 2023.

\bibitem{mu2024automatic}
B.~Mu, P.~Guo, D.~Guo, P.~Zhou, W.~Chen, and L.~Xie, ``Automatic channel selection and spatial feature integration for multi-channel speech recognition across various array topologies,'' in \emph{Proc. of IEEE ICASSP}, 2024.

\bibitem{kanda2020serialized}
N.~Kanda, Y.~Gaur, X.~Wang, Z.~Meng, and T.~Yoshioka, ``Serialized output training for end-to-end overlapped speech recognition,'' in \emph{Proc. of Interspeech}, 2020.

\bibitem{carletta2005ami}
J.~Carletta, S.~Ashby, S.~Bourban, M.~Flynn, M.~Guillemot, T.~Hain, J.~Kadlec, V.~Karaiskos, W.~Kraaij, M.~Kronenthal \emph{et~al.}, ``The {AMI} meeting corpus: A pre-announcement,'' in \emph{Proc. of MLMI}, 2005.

\bibitem{vincent2018audio}
E.~Vincent, T.~Virtanen, and S.~Gannot, \emph{Audio Source Separation and Speech Enhancement}.\hskip 1em plus 0.5em minus 0.4em\relax John Wiley \& Sons, 2018.

\bibitem{don1993array}
D.~H. Johnson and D.~E. Dudgeon, \emph{{Array Signal Processing: Concepts and Techniques}}.\hskip 1em plus 0.5em minus 0.4em\relax Prentice Hall., 1993.

\bibitem{yang2023simulating}
M.~Yang, N.~Kanda, X.~Wang, J.~Wu, S.~Sivasankaran, Z.~Chen, J.~Li, and T.~Yoshioka, ``Simulating realistic speech overlaps improves multi-talker {ASR},'' in \emph{Proc. of IEEE ICASSP}, 2023.

\bibitem{panayotov2015LibriSpeech}
V.~Panayotov, G.~Chen, D.~Povey, and S.~Khudanpur, ``Librispeech: An {ASR} corpus based on public domain audio books,'' in \emph{Proc. of IEEE ICASSP}, 2015.

\bibitem{kanda2020joint}
N.~Kanda, Y.~Gaur, X.~Wang, Z.~Meng, Z.~Chen, T.~Zhou, and T.~Yoshioka, ``Joint speaker counting, speech recognition, and speaker identification for overlapped speech of any number of speakers,'' in \emph{Proc. of Interspeech}, 2020.

\bibitem{kudo2018sentencepiece}
T.~Kudo and J.~Richardson, ``Sentencepiece: A simple and language independent subword tokenizer and detokenizer for neural text processing,'' in \emph{Proc. of EMNLP}, 2018.

\bibitem{lu2022towards}
Y.-J. Lu, S.~Cornell, X.~Chang, W.~Zhang, C.~Li, Z.~Ni, Z.-Q. Wang, and S.~Watanabe, ``Towards low-distortion multi-channel speech enhancement: The {ESPNet-SE submission to the L3DAS22} challenge,'' in \emph{Proc. of IEEE ICASSP}, 2022.

\bibitem{kim23b_interspeech}
M.~Kim, S.~Cheong, and J.~W. Shin, ``{DNN-based Parameter Estimation for MVDR Beamforming and Post-filtering},'' in \emph{Proc. of Interspeech}, 2023.

\bibitem{yang2021torchaudio}
Y.-Y. Yang, M.~Hira, Z.~Ni, A.~Chourdia, A.~Astafurov, C.~Chen, C.-F. Yeh, C.~Puhrsch, D.~Pollack, D.~Genzel, D.~Greenberg, E.~Z. Yang, J.~Lian, J.~Mahadeokar, J.~Hwang, J.~Chen, P.~Goldsborough, P.~Roy, S.~Narenthiran, S.~Watanabe, S.~Chintala, V.~Quenneville-Bélair, and Y.~Shi, ``Torchaudio: Building blocks for audio and speech processing,'' \emph{arXiv preprint arXiv:2110.15018}, 2021.

\bibitem{speechbrain}
M.~Ravanelli, T.~Parcollet, P.~Plantinga, A.~Rouhe, S.~Cornell, L.~Lugosch, C.~Subakan, N.~Dawalatabad, A.~Heba \emph{et~al.}, ``{SpeechBrain}: A general-purpose speech toolkit,'' 2021, arXiv:2106.04624.

\end{thebibliography}
\end{document}